# Noncollinear electro-optic sampling detection of terahertz pulses in a $LiNbO_3$ crystal with avoiding the effect of intrinsic birefringence

A. I. SHUGUROV,[1] S. B. BODROV,[1,2] E. A. MASHKOVICH,[1,3] H. KITAHARA,[4] N. A. ABRAMOVSKY,[1] M. TANI,[4] AND M. I. BAKUNOV[1,*]

[1]*University of Nizhny Novgorod, Nizhny Novgorod, Russia*
[2]*Institute of Applied Physics, Russian Academy of Sciences, Nizhny Novgorod, Russia*
[3]*Radboud University, Institute for Molecules and Materials, 6525 AJ Nijmegen, the Netherlands*
[4]*Research Center for Development of Far-Infrared Region, University of Fukui, Fukui, Japan*
**bakunov@rf.unn.ru*

**Abstract:** We propose and experimentally prove efficient high-resolution electro-optic sampling measurement of broadband terahertz waveforms in a $LiNbO_3$ crystal in the configuration with the probe laser beam propagating along the optical axis of the crystal. This configuration allows one to avoid the detrimental effect of strong intrinsic birefringence of $LiNbO_3$ without any additional optical elements. To achieve velocity matching of the terahertz wave and the probe beam, the terahertz wave is introduced into the crystal through a Si prism at the Cherenkov angle to the probe beam. The workability of the scheme at different wavelengths of the probe optical beam (800 and 1550 nm) is demonstrated.

## 1. Introduction

Free space electro-optic (EO) sampling of terahertz waveforms by femtosecond laser pulses is widely used to characterize broadband terahertz pulses, especially in terahertz time-domain spectroscopy. In the conventional EO sampling scheme, the probe optical pulse propagates collinearly with the terahertz pulse in an EO crystal and acquires a terahertz-field-induced polarization change via the Pockels effect. By measuring the polarization change as a function of the time delay between the pulses, one can map the time-dependence of the terahertz electric field [1]. Efficient EO sampling requires velocity matching between optical group and terahertz phase velocities. For a given wavelength $\lambda$ of the probe optical beam, this requirement can be achieved only in a specific crystal, for example, in ZnTe for a Ti:sapphire laser with $\lambda \approx 800$ nm [2].

Recently, a noncollinear EO sampling scheme, where the probe optical beam propagates at the Cherenkov angle to the terahertz beam, was proposed [3] as a way to achieve optical-terahertz velocity matching even in crystals with a large collinear velocity mismatch, such as $LiNbO_3$. In the experimental demonstration of the technique [3], taking into account a large (~63° [4]) Cherenkov angle in $LiNbO_3$, terahertz radiation was introduced into the crystal through its lateral surface by using a Si-prism coupler. To reduce substantial terahertz absorption in $LiNbO_3$, the probe optical beam was adjusted parallel and close to the crystal-prism interface.

The most salient advantage of the noncollinear (Cherenkov-type) EO sampling scheme with a $LiNbO_3$ crystal is its workability at different wavelengths of the probe optical beam. Also conveniently, the weak optical dispersion of $LiNbO_3$ [4] allows to use a Si-prism coupler with the same cut angle in schemes with Ti:sapphire ($\lambda \approx 800$ nm), Yb-doped ($\lambda \approx 1.06$ $\mu$m), and fiber ($\lambda \approx 1.55$ $\mu$m) lasers. For comparison, using GaAs, instead of $LiNbO_3$, together with a fiber laser ($\lambda \approx 1.55$ $\mu$m) can be even more convenient due to a smaller (~12°) Cherenkov angle (this allows to omit the Si-prism coupler) and smaller terahertz absorption [5]. GaAs cannot, however, be used with the most common Ti:sapphire laser because of the crystal's opacity. The general advantage of the noncollinear EO sampling schemes with both $LiNbO_3$ and GaAs crystals is their capability to operate with cm-thick crystals. This allows one to use wide time windows of EO sampling and, therefore, to achieve high (about a few GHz) spectral resolution of terahertz detection.

In the scheme introduced in Ref. [3], the *z*-axis of the $LiNbO_3$ crystal is oriented perpendicularly to the propagation directions of the probe and terahertz beams. The terahertz beam is polarized along the *z*-axis, whereas the polarization of the probe beam is at 45° to the *z*-axis. Such configuration allows to exploit the largest electro-optic coefficient of $LiNbO_3$ $r_{33} \approx 31$ pm/V [6]. A disadvantage of the configuration, however, is a parasitic effect of strong intrinsic birefringence of $LiNbO_3$, with group refractive indices 2.3 of ordinary and 2.2 of extraordinary waves at the wavelength of 800 nm [7]. After propagation of the probe optical pulse with a typical 100-fs duration through a 0.3-mm thick $LiNbO_3$ crystal the orthogonally polarized components of the probe pulse will be spatially separated thus preventing the ellipsometric measurements. To compensate the effect of the intrinsic birefringence, special efforts, such as proposed in Ref. [8], are required. This substantially complicates the experimental scheme [3].

In this paper, we propose and experimentally demonstrate a noncollinear EO sampling scheme with a $LiNbO_3$ crystal, whose *z*-axis is oriented along the propagation direction of the probe optical pulse. In this degenerate configuration, the probe optical pulse of any polarization propagates in the crystal as an ordinary wave not experiencing a parasitic effect of the intrinsic birefringence of $LiNbO_3$. A change of the probe beam polarization is only determined by the terahertz-electric-field induced birefringence. Although in this configuration the ellipsometric signal is proportional to the electro-optic coefficient $r_{22}$ [9], which is not as large as $r_{33}$, the experimentally achieved dynamic range is comparable to those obtained by other methods, such as EO sampling in GaAs and photoconductive sampling. At the same time, the proposed technique retains the advantages of high spectral resolution and workability at different optical wavelengths.

## 2. Experimental setup

Experiments were performed with a structure consisting of a 2-mm thick and 1×1 cm2 in size plate of LiNbO3 clamped to a trapezoidal Si prism [Fig. 1(a)]. The prism is cut at the angle 41° providing the Cherenkov synchronization of a terahertz pulse with the probe optical pulse propagating in the LiNbO3 plate along the LiNbO3-Si interface and the crystal z-axis. The polarizations of both the terahertz and probe beams were set along the crystal x-axis.

In the first experimental configuration [Fig. 1(b)], a femtosecond $Er^{3+}$–doped fiber laser (1.55 μm central wavelength, 70 fs pulse duration, and 100 MHz repetition rate) was used as a light source for terahertz generation and detection. The laser beam was split into pump and probe beams. The pump beam (35 mW average power) triggered a photoconductive antenna (PCA) on a InGaAs/InAlAs substrate, which was biased with a ±20 V, 10 kHz square wave voltage. The generated terahertz radiation was collimated by a TPX lens and focused by a parabolic mirror through a Si prism onto the interface between the prism and $LiNbO_3$ plate. From knife-edge measurements with use of a Golay cell, the focal spot diameter (FWHM) was estimated to be about 2.2 mm. PCA and TPX lens were placed on a translational stage to vary the arrival time between the terahertz pulse and probe pulse on the $LiNbO_3$ plate.

The probe beam (30 mW) was collimated by a 25.4 mm focal length lens $f_1$ and focused onto the 0.2×1 $cm^2$ entrance surface of the $LiNbO_3$ plate by a lens $f_2$ with 200 mm focal length [Fig. 1(b)]. The 1/*e* focal spot diameter was about 90 $\mu$m. The probe beam polarization was cleaned up with a Glan prism (GP). A common combination of a quarter-wave plate ($\lambda$/4), Wollaston prism (WP), and balanced photodetector was used to measure the terahertz-induced ellipticity of the probe beam. To improve the detection of high terahertz frequencies, which suffer substantial absorption while propagating into the $LiNbO_3$ plate from the $LiNbO_3$-Si interface, we used an iris. The iris was adjusted to transmit only the part of the probe beam that propagated in the vicinity of the $LiNbO_3$-Si interface thus increasing the EO signal at the high frequencies.

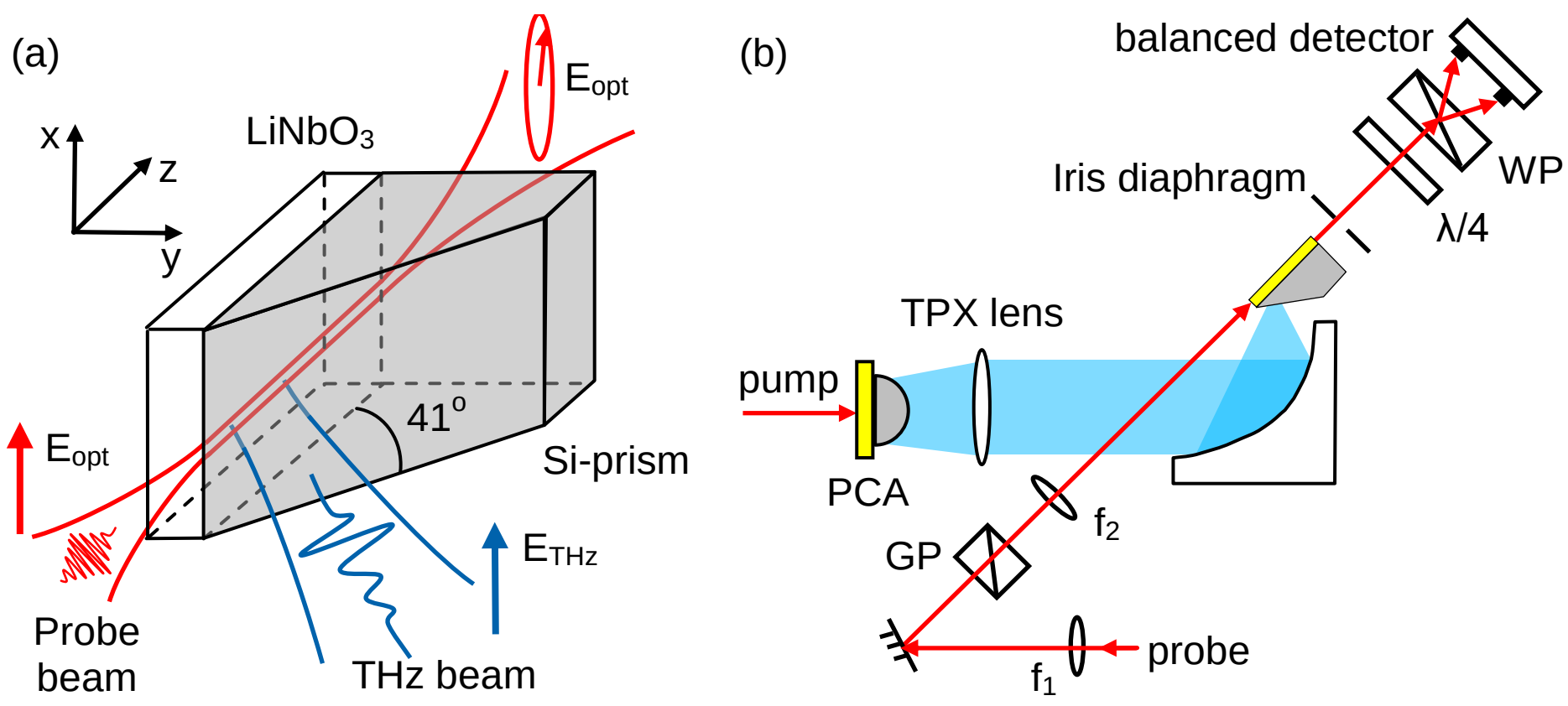

Fig. 1. (a) Geometry of noncollinear propagation of the terahertz (blue) and probe optical (red) beams in the $LiNbO_3$-Si-prism structure. (b) Schematic of the experimental setup with a fiber laser ($\lambda$ = 1.55 μm) as light source and PCA as a terahertz source.

In the second experimental configuration (Fig. 2), a Ti:sapphire laser (800 nm central wavelength, 80 fs pulse duration, and 80 MHz repetition rate) was used as a light source. Terahertz radiation was generated in a Cherenkov-type optical-to-terahertz converter, which comprised a 55-$\mu$m thick layer of $LiNbO_3$ clamped between two Si prisms of total internal reflection for terahertz radiation [10]. Emitted from the converter terahertz radiation was collected, collimated, and focused to the $LiNbO_3$-Si-prism detection structure [Fig. 1(a)] by a pair of parabolic mirrors. Thus, the second experimental setup (Fig. 2) was fully $LiNbO_3$-based and fully Cherenkov-type. The 1/*e* probe beam diameter at the entrance facet of the $LiNbO_3$ detector crystal was about 40 $\mu$m.

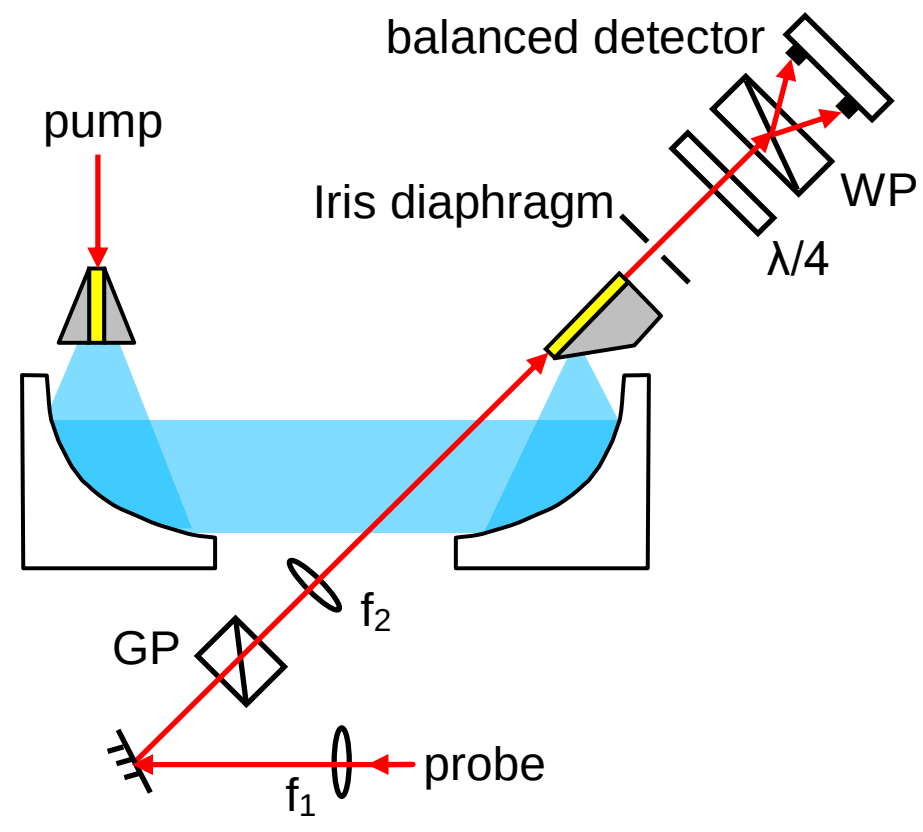

Fig. 2. Schematic of the experimental setup with a Ti:sapphire laser ($\lambda$ = 800 nm) as a light source and Cherenkov-type $LiNbO_3$-based optical-to-terahertz converter as a terahertz source.

## 3. Theoretical analysis

First of all, let us justify the choice of the prism apex angle. The probe optical pulse propagates in $LiNbO_3$ as an ordinary wave with a group velocity $c/n_g$, where $c$ is the speed of light and $n_g$ is the ordinary group refractive index. The wavefronts of the terahertz wave propagate in Si with the velocity $c/n_{Si}$. The intersection point of a wavefront with the Si-$LiNbO_3$ interface moves at a velocity $c/(n_{Si}\sin\alpha)$, where $\alpha$ is the angle between the wavefront and the interface. In the $LiNbO_3$ slab, the intersection point of the wavefront with the probe beam axis evidently moves with the same velocity. By equating this velocity to the probe pulse velocity $c/n_g$, one can obtain $\sin\alpha = n_g/n_{Si}$. Substitution of $n_{Si} = 3.42$ [11] and $n_g = 2.26$ (as calculated from the Sellmeier equation for an ordinary wave at 1.55 $\mu$m [12]) yields $\alpha \approx 41°$. Since terahertz wavefronts in the Si prism are parallel to the prism entrance face, the prism apex angle equals $\alpha$, i.e., 41°. For this angle, due to a flat refractive index of high-resistivity Si at terahertz frequencies, the optical pulse and terahertz wave are almost perfectly synchronized in a wide terahertz frequency range. For the probe beam wavelength of 800 nm, $n_g = 2.35$ [12] and $\alpha \approx 43°$. In practice, due to the small difference in $\alpha$, the same prism can be used for both wavelengths.

Now let us justify the choice of the probe beam polarization. For the terahertz electric field $\mathbf{E}^{\mathrm{THz}}$ applied in the $x$-direction and the optical wave propagating in the $z$-direction, the transverse (with respect to the $z$-axis) impermeability tensor of $LiNbO_3$ is given by [13]

$$\eta_{ij} = \begin{pmatrix} n_o^{-2} & -r_{22}E^{\mathrm{THz}} \\ -r_{22}E^{\mathrm{THz}} & n_o^{-2} \end{pmatrix}, \tag{1}$$

where $n_o$ is the ordinary optical refractive index and $r_{22}$ is a component of the electro-optic tensor. To diagonalize the tensor, we find its eigenvalues $\Lambda_{1,2} = n_o^{-2} \mp r_{22}E^{\mathrm{THz}}$ and corresponding principal axes $x' = x + y$, $y' = y - x$. The principal axes $x', y'$ are rotated by 45° with respect to the $x, y$ axes. Thus, the largest polarization change occurs for the probe beam polarized either in the $x$-direction (as in our setup) or in the $y$-direction.

The principal values of the refractive index are given by $n_{x\prime} = \Lambda_1^{-1/2} \approx n_o(1 + n_o^2 r_{22}E^{\mathrm{THz}}/2)$, $n_{y\prime} = \Lambda_2^{-1/2} \approx n_o(1 - n_o^2 r_{22}E^{\mathrm{THz}}/2)$. For perfect velocity matching, which is readily accessible with the $LiNbO_3$-Si-prism structure [Fig. 1(a)] in a wide terahertz frequency range, the EO signal $\Delta I/I$ is given by the relative phase retardation $\Delta\varphi$ between the probe beam polarization components along the $x'$ and $y'$ axes

$$\Delta I/I = \Delta\varphi = \frac{2\pi}{\lambda} n_o^3 r_{22} E^{\mathrm{THz}} L, \tag{2}$$

where $\lambda$ is the laser wavelength and $L$ is the interaction length. The length $L$ is defined by the size of the terahertz beam spot on the Si-$LiNbO_3$ interface in the $z$-direction, i.e., $L \sim D^{\mathrm{THz}}/\cos 41° \sim 3$ mm (with $D^{\mathrm{THz}} \sim 2.2$ mm the terahertz beam width). The crystal parameters are $n_o = 2.23$ [7,12] and $r_{22} \approx 4$ pm/V [14]. Although $r_{22}$ is several times smaller than $r_{33} \approx 31$ pm/V, it is sufficient for a reliable detection (see Sec.4).

Importantly, the optical-terahertz interaction length $L \sim 3$ mm in the proposed Cherenkov-type scheme is ~30 times larger than the optical-terahertz coherence length in $LiNbO_3$ in the collinear configuration [3,9]. According to Eq. (2), this increases substantially the EO signal. Additionally, the lateral input of terahertz radiation into the $LiNbO_3$ layer in the Cherenkov-type scheme allows for diminishing the negative effect of strong terahertz absorption in $LiNbO_3$, which affects significantly the EO response in the collinear geometry [9]. The effects of terahertz absorption and velocity mismatch can be reduced in the collinear geometry by using thin $LiNbO_3$ crystals. This, however, leads to the appearance of echo signals and, as a result, to a modulation of the terahertz spectrum or to a loss of spectral resolution if the echo signals are filtered out by shrinking the time window. In the proposed $LiNbO_3$-Si-prism structure, the $LiNbO_3$ layer is 1 cm long and, therefore, the spectral resolution can be as high as a few GHz.

## 4. Results and discussion

Figure 3 shows the terahertz spectrum obtained by means of EO sampling in the proposed $LiNbO_3$-Si-prism structure in the experimental setup with a fiber laser [Fig. 1(b)]. The spectra of the terahertz signal from the same PCA obtained by photoconductive sampling in another PCA and by noncollinear EO sampling in a 4-mm thick GaAs crystal [5] are shown for reference. From Fig. 3, it is seen that the efficiencies of the three methods are almost the same. However, the spectral resolution achieved with the $LiNbO_3$-Si-prism detection structure (about 10 GHz) is substantially higher than the resolution (about 46 GHz) provided by the fully PCA-based terahertz spectrometer. The $LiNbO_3$-Si-prism structure shows a smaller response at the high frequencies than EO sampling in a GaAs crystal. This can be attributed to a higher terahertz absorption in $LiNbO_3$ than in GaAs. GaAs, however, cannot be used with a Ti:sapphire laser.

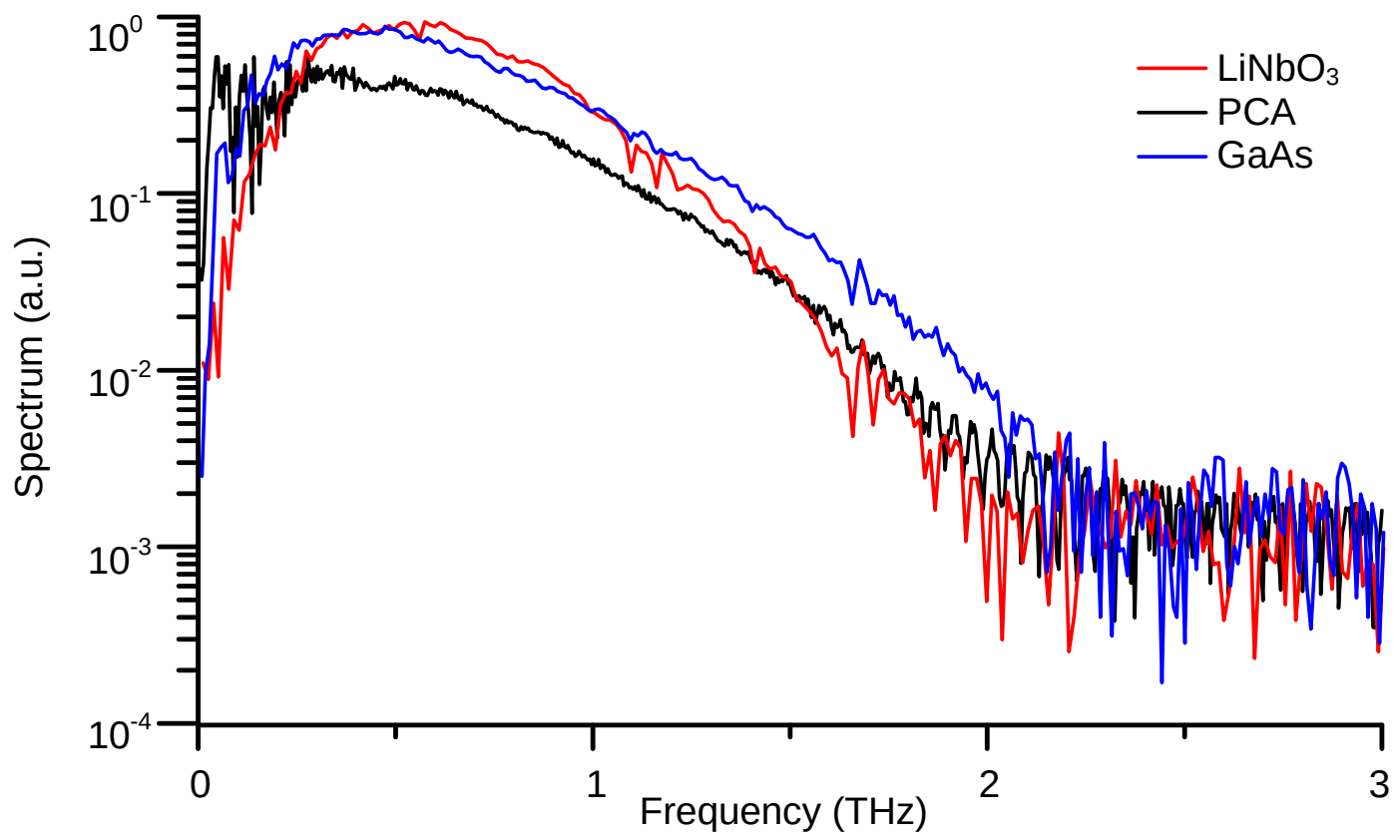

Fig. 3. Terahertz spectrum measured with the $LiNbO_3$-Si-prism structure in the setup with a fiber laser (red). The spectra obtained in the fully PCA-based setup (black) and by EO sampling in a 4-mm thick GaAs crystal (blue) are shown for reference.

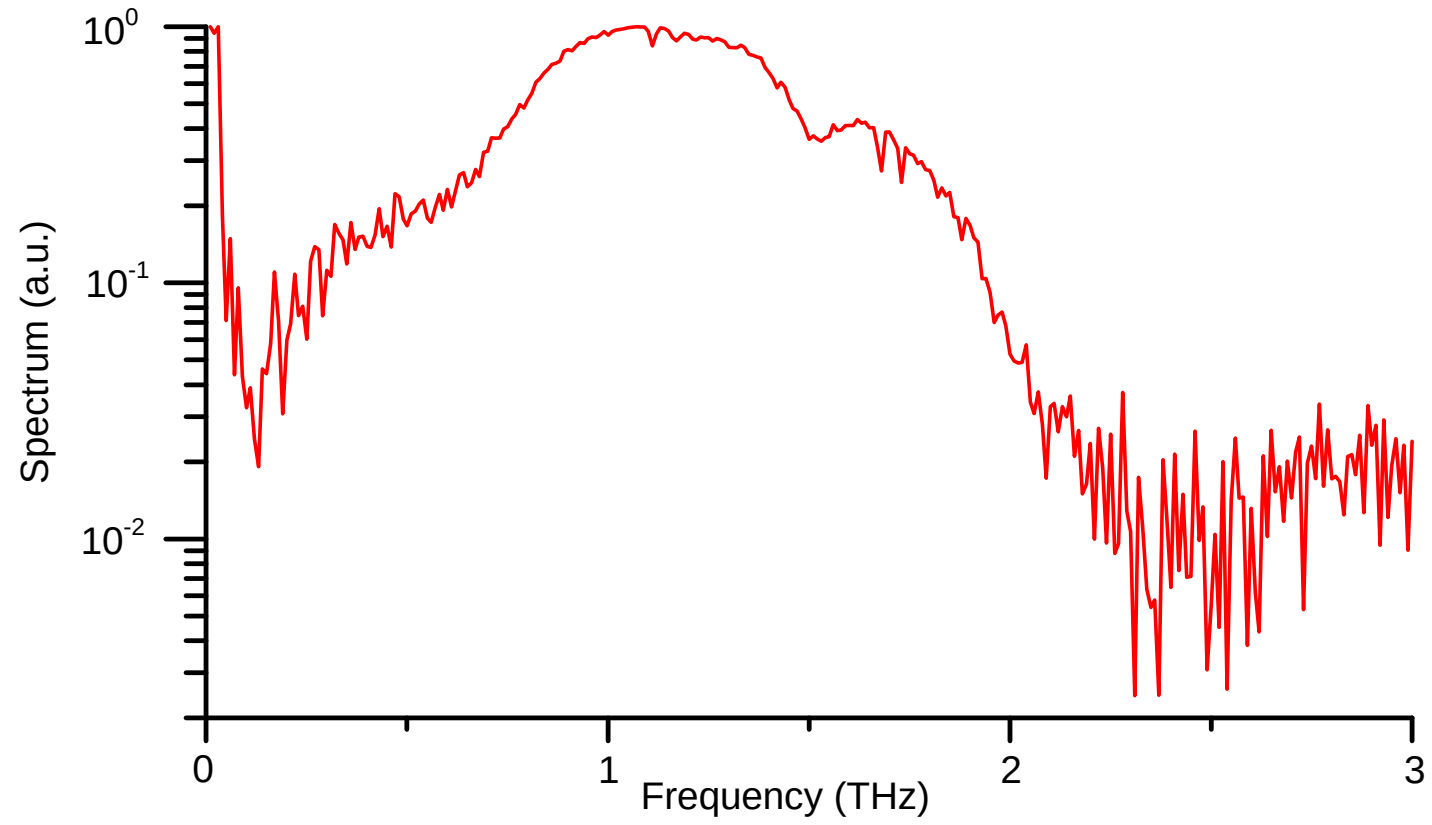

Fig. 4. Terahertz spectrum measured with the $LiNbO_3$-Si-prism structure in the setup with a Ti:sapphire laser.

Figure 4 shows the terahertz spectrum obtained with the proposed $LiNbO_3$-Si-prism structure in the setup with a Ti:sapphire laser (Fig. 2), i.e., in a fully $LiNbO_3$-based and fully Cherenkov-type setup. In this setup, optical rectification of femtosecond laser pulses in the $LiNbO_3$-based Cherenkov-type optical-to-terahertz converter produces terahertz radiation with higher frequencies, as compared to the PCA emitter. In particular, the spectrum maximum is at about 1.5 THz [10]. This agrees well with the spectrum in Fig. 4. More generally, the spectrum shape in Fig. 4 is close to that in Ref. [10] for all frequencies up to about 2 THz. The frequencies above 2 THz are not detected in our setup, although in Ref. [10], where a 1-mm thick ZnTe crystal is used as detector, the spectrum continues up to almost 3 THz. The bandwidth limitation can be attributed to a finite width (≈40 $\mu$m) of the probe optical beam and the noncollinear geometry of the optical-terahertz interaction [5, 15, 16]. The detection bandwidth can be increased by reducing the thickness of the $LiNbO_3$ layer [16]. The spectral resolution in Fig. 4 (<10 GHz) is much higher than the resolution (>40 GHz) provided by a 1-mm thick ZnTe detector crystal [10].

## 5. Summary

The proposed Cherenkov-type $LiNbO_3$-Si-prism structure can be used as a universal tool for EO sampling of terahertz radiation at different wavelengths of the probe optical beam, in particular, with convenient fiber ($\lambda \approx 1.55$ $\mu$m) and widespread Ti:sapphire ($\lambda \approx 800$ nm) lasers as a light source. Due to the propagation of the probe optical beam along the *z*-axis of the $LiNbO_3$ crystal, the structure allows for avoiding the effect of strong intrinsic birefringence in $LiNbO_3$ without using any additional optical elements. The long (~1 cm) propagation distance of the probe beam in the $LiNbO_3$ layer ensures high spectral resolution (<10 GHz) of terahertz detection with the proposed structure.

**Funding.** Ministry of Science and Higher Education of the Russian Federation (FSWR-2021-011), Russian Foundation for Basic Research (20-32-90080).

**Disclosures.** The authors declare no conflicts of interest.

**Data availability.** Data underlying the results presented in this paper are not publicly available at this time but may be obtained from the authors upon reasonable request.

## References

1. Q. Wu and X.-C. Zhang, "Ultrafast electro-optic field sensors," Appl. Phys. Lett. **68**(12), 1604–1606 (1996).
2. M. Nagai, K. Tanaka, H. Ohtake, T. Bessho, T. Sugiura, T. Hirosumi, and M. Yoshida, "Generation and detection of terahertz radiation by electro-optical process in GaAs using 1.56 $\mu$m fiber laser pulses," Appl. Phys. Lett. **85**(18), 3974–3976 (2004).
3. M. Tani, K. Horita, T. Kinoshita, C. T. Que, E. Estacio, K. Yamamoto, and M. I. Bakunov, "Efficient electro-optic sampling detection of terahertz radiation via Cherenkov phase matching," Opt. Express. **19**(21), 19901–19906 (2011).
4. J. A. Fülöp, L. Pálfalvi, G. Almási, and J. Hebling, "Design of high-energy terahertz sources based on optical rectification," Opt. Express **18**(12), 12311–12327 (2010).
5. E. A. Mashkovich, A. I. Shugurov, S. Ozawa, E. Estacio, M. Tani, and M. I. Bakunov, "Noncollinear electro-optic sampling of terahertz waves in a thick GaAs crystal," IEEE Trans. Terahertz Sci. Technol. **5**(5), 732–736 (2015).
6. R. W. Boyd, *Nonlinear Optics*, 3rd ed. (Academic, 2008).
7. E. A. Mashkovich, S. A. Sychugin, and M. I. Bakunov, "Generation of narrowband terahertz radiation by an ultrashort laser pulse in a bulk $LiNbO_3$ crystal," J. Opt. Soc. Am. B **34**(9), 1805–1810 (2017).
8. P. Y. Han, M. Tani, F. Pan, and X.-C. Zhang, "Use of the organic crystal DAST for terahertz beam applications," Opt. Lett. **25**(9), 675–677 (2000).
9. C. Winnewisser, P. U. Jepsen, M. Schall, V. Schyja, and H. Helm, "Electro-optic detection of THz radiation in $LiTaO_3$, $LiNbO_3$ and ZnTe," Appl. Phys. Lett. **70**(23), 3069–3071 (1997).
10. M. I. Bakunov, E. S. Efimenko, S. D. Gorelov, N. A. Abramovsky, and S. B. Bodrov, "Efficient Cherenkov-type optical-to-terahertz converter with terahertz beam combining," Opt. Lett. **45**(13), 3533–3536 (2020).
11. D. Grischkowsky, S. Keiding, M. van Exter, and C. Fattinger, "Far-infrared time-domain spectroscopy with terahertz beams of dielectrics and semiconductors," J. Opt. Soc. Am. B **7**(10), 2006–2015 (1990).
12. D. E. Zelmon, D. L. Small, and D. Jundt, "Infrared corrected Sellmeier coefficients for congruently grown lithium niobite and 5 mol. % magnesium oxide-doped lithium niobate," J. Opt. Soc. Am. B **14**(12), 3319–3322 (1997).

13. A. Yariv and P. Yeh, *Optical Waves in Crystals* (Wiley, 1984).
14. G. E. Jellison Jr., C. O. Griffiths, and D. E. Holcomb, "Electric-field-induced birefringence in $LiNbO_3$ measured by generalized transmission ellipsometry," Appl. Phys. Lett. **81**(7), 1222–1224 (2002).
15. M. I. Bakunov, S. D. Gorelov, and M. Tani, "Nonellipsometric noncollinear electrooptic sampling of terahertz waves: a comprehensive theory," IEEE Trans. Terahertz Sci. Technol. **6**(3), 473–479 (2016).
16. I. E. Ilyakov, B. V. Shishkin, S. B. Bodrov, G. K. Kitaeva, M. I. Bakunov, and R. A. Akhmedzhanov, "Highly sensitive electro-optic detection of terahertz waves in a prism-coupled thin $LiNbO_3$ layer," Laser Phys. Lett. **17**, 085403 (2020).